\documentclass[
reprint,
 amsmath,amssymb,
 aip,showkeys,jap
]{revtex4-1}

\usepackage{enumerate}
\usepackage{color}
\usepackage{float}
\usepackage{wrapfig}
\usepackage{verbatim}
 \usepackage{braket}
\usepackage{xcolor}

\newcommand\hl[1]{
  \bgroup
  \hskip0pt\color{red!80!black}
  #1
  \egroup
}

\usepackage[ruled,vlined,linesnumbered]{algorithm2e}

\def\R{\mathrm{R}}

\usepackage{graphicx}% Include figure files
\usepackage{dcolumn}% Align table columns on decimal point
\usepackage{bm}% bold math

\usepackage[utf8]{inputenc}
\usepackage[T1]{fontenc}
\usepackage{mathptmx}
\usepackage{etoolbox}

\makeatletter
\def\@email#1#2{
 \endgroup
 \patchcmd{\titleblock@produce}
  {\frontmatter@RRAPformat}
  {\frontmatter@RRAPformat{\produce@RRAP{*#1\href{mailto:#2}{#2}}}\frontmatter@RRAPformat}
  {}{}
}
\makeatother

\begin{document}

\preprint{AIP/123-QED}

\title{High-precision real-space simulation of \\ electrostatically-confined few-electron states}% Force line breaks with \\

\author{Christopher R. Anderson}
\affiliation{ 
Department of Mathematics, University of California, Los Angeles, Los Angeles, CA 90095 USA}
\affiliation{
Center for Quantum Science and Engineering, University of California, Los Angeles, Los Angeles, CA 90095 USA
}

\author{Mark F. Gyure}
\affiliation{
Center for Quantum Science and Engineering, University of California, Los Angeles, Los Angeles, CA 90095 USA
}
\affiliation{HRL Laboratories, LLC, 3011 Malibu Canyon Road, Malibu, California 90265, USA}

\author{Sam Quinn}
\affiliation{HRL Laboratories, LLC, 3011 Malibu Canyon Road, Malibu, California 90265, USA}

\author{Andrew Pan}
\affiliation{HRL Laboratories, LLC, 3011 Malibu Canyon Road, Malibu, California 90265, USA}

\author{Richard S. Ross}
\altaffiliation[Current address: ]{Department of Physics, University of California, Los Angeles, Los Angeles, CA 90095 USA}

\author{Andrey A. Kiselev}
\affiliation{HRL Laboratories, LLC, 3011 Malibu Canyon Road, Malibu, California 90265, USA}

\date{\today}

\begin{abstract}
In this paper we present a computational procedure that utilizes real-space grids to obtain high precision approximations of electrostatically confined few-electron states such as those that arise in gated semiconductor quantum dots. We use the Full Configuration Interaction (FCI) method with a continuously adapted orthonormal orbital basis to approximate the ground and excited states of such systems. We also introduce a benchmark problem based on a realistic analytical electrostatic potential for quantum dot devices.  We show that our approach leads to highly precise computed energies and energy differences over a wide range of model parameters. The analytic definition of the benchmark allows for a collection of tests that are easily replicated, thus facilitating comparisons with other computational approaches.  
\end{abstract}


\maketitle

\pagestyle{myheadings} \thispagestyle{plain}
\renewcommand\thesection{\arabic{section}}
\renewcommand\thesubsection{\thesection.\arabic{subsection}}

\section{Introduction}

Semiconductor spin qubits show great promise for quantum information applications. \cite{Hanson2007,Zwanenberg2013, burkard_semiconductor_2021} They rely on precise control of the spatial and spin structure of few-electron quantum dot states. The quantum dots are often realized in planar semiconductor heterostructures where the electrons are confined out-of-plane by a quantum well or metal-oxide-semiconductor (MOS) interface, and in-plane by the electrostatic potential due to voltage-biased gate electrodes. In single- or multi-qubit devices, information is contained within quantum dot states whose relevant energy splittings can span a wide range from sub-neV to meV. In particular, a crucial ingredient for many types of spin qubits is the Heisenberg exchange interaction, $J$, between nearby electrons. Exchange can be used for generating two-qubit gates between individual spins or for controlling a single encoded qubit defined in a multi-spin system. In the absence of an external magnetic field, the exchange energy for two electrons is equal to the energy gap between the ground (singlet) and first excited (triplet) states and can be tuned over many orders of magnitude by changes in the electrostatic potential induced by gate voltage adjustments. \cite{Reed2016}

High-precision calculations of the structure of low-energy levels are therefore necessary to simulate the physics and improve understanding and design of spin qubit devices. This requires finding convergent solutions of the $N$-particle Schr{\"o}dinger equation, accounting for realistic electrostatic confinement and the electron-electron Coulomb interaction. The full configuration interaction (FCI) method is a powerful approach for solving this problem, capable of providing accurate numerical solutions for the ground and excited states of such systems. First developed in the context of quantum chemistry, it has been increasingly adopted for analyzing quantum dots, \cite{ friesen_practical_2003,Rontani2006} donors, \cite{Tankasala2018,Joecker2021} and other artificial few-electron systems, \cite{Reimann2002} enabling illuminating studies into their level structure, \cite{ercan_strong_2021, abadillo-uriel_two-body_2021} susceptibility to charge noise \cite{nielsen_implications_2010, bakker_validity_2015}  and disorder, \cite{rahman_voltage_2012} relaxation rates, \cite{climente_effect_2006} and operating regimes for mediated exchange \cite{deng_interplay_2020} and microwave coupling. \cite{pan_resonant_2020}
The FCI method constructs approximate $N$-particle solutions from a finite set of orbital basis functions. The number and quality of these orbital basis functions should be chosen well for precise, computationally efficient, and physically representative simulations. Many quantum dot calculations use a set of analytical basis functions and simple confinement
models like parabolic or quartic potentials. However, because quantum dot states are strongly dependent on the confining potential, realistic electrostatic potentials, typically obtained as numerical solutions of Poisson's equation, should preferentially be used along with basis functions that reflect the key features of those potentials.

In this work, we present an approach to FCI calculations tailored for simulating realistic quantum dot devices with high
precision. In particular, it enables an accurate tracking of the exchange over many orders of magnitude as gate biases are
varied — a much more stringent test of simulation capabilities than requiring a high precision value of $J$ for a single
bias configuration. We use a real space grid approximation framework to accommodate arbitrary confining potentials. 
Special consideration is given to the construction of adaptive, orthonormal orbital basis functions and the 
evaluation of one- and two-particle integrals on those grids. 

In order to validate our computational procedure we present results on a benchmark problem that captures the full range
of realistic quantum dot operation, ranging from situations of weakly interacting electrons, relatively isolated in individual
quantum dots, to cases where they all may be confined in a single quantum dot. This benchmark problem is a good 
approximation to the potentials associated with the modulation of states in a double quantum dot configuration. A key feature
of this benchmark problem is that the potential is a parameterized analytic function, an aspect that greatly facilitates assessing the accuracy of the FCI simulations over a range of realistic conditions. 

In the next section we describe the general FCI procedure for computing solutions to the $N$-particle Schr{\"o}dinger equation and present the details of our grid-based implementation of its steps. The benchmark problem is described in the third section, and in the fourth section we present the results obtained using the grid-based computational procedure for this benchmark problem.  

\section{Grid based Full Configuration Interaction (FCI) Procedure}

\bigskip
The computational procedures presented in this paper are described in the context of a completely grid based computational method for determining the eigenfunctions and eigenvalues of a few-electron Hamiltonian associated with the effective mass approximation for electron states within a semiconductor. This problem takes the form $\mathrm{H}  \Psi = \lambda \Psi$ where the Hamiltonian, $\mathrm{H}$, is the linear operator
\begin{equation}
\mathrm{H} \, = \, 
\sum\limits_{i=1}^{N}
\Big[ \underbrace{ -\frac{\hbar^2}{2} \nabla_i \cdot \bm{\beta} \cdot \nabla_i  + \mathrm{U} (\bm{r}_{i}) }_{\mathrm{h}_i} \Big]
+ \sum_{i=1}^{N} \sum_{j > i}^{N} \frac{\alpha}{{| \bm{r}_i - \bm{r}_j |}}  
\label{NparticleS}
\end{equation}
and $\Psi = \Psi(\tilde r_1,\, \tilde r_2, \, \ldots \, \tilde r_N)$ is the $N$-particle eigenfunction; $\tilde r_i = (\bm{r}_i, s_i)$ 
with $\bm{r}_i \in \mathrm{R}^3$ being the spatial coordinate and $s_i$ being the spin coordinate of the $i$th particle. The external potential $\mathrm{U}$ and the reciprocal effective mass tensor $\bm{\beta}$, chosen with $\mathrm{diag}\,\bm{\beta} = (m_{x}^{-1}, m_{y}^{-1}, m_{z}^{-1} )$, define the single-particle Hamiltonian $\mathrm{h}$ (we ignore magnetic fields and spin-orbit effects in this paper, making $\mathrm{h}$ and $\mathrm{H}$ spin independent). In SI notation, $\alpha =  e^2/4\pi\epsilon$ where $e$ is the electron charge and $\epsilon$ is the absolute macroscopic static dielectric constant of the semiconductor material. The effective mass model and coordinate-independent values for $\bm{\beta}$ and $\varepsilon$ suffice for describing the main properties of interest in this paper. However, we note that the FCI methodology can be extended beyond the effective mass approximation to use a multi-band model such as $\bm{k}\cdot\bm{p}$ or tight-binding Hamiltonians, which can be important for simulating valence band holes, valley and spin-orbit physics in silicon devices, or other band structure effects. \cite{ercan_strong_2021,nielsen_many-electron_2012,secchi_interacting_2021} Similarly, coordinate-dependent $\bm{\beta}$ and $\varepsilon$ values, usually present in semiconductor heterostructures, can also be accommodated in our approach with proper attention to underlying subtleties. \footnote{With spatially varying effective mass, order of operators in the kinetic term of Eq.~(\ref{NparticleS}) is important. As written, it does preserve Hamiltonian hermiticity, but the formulation is actually not unique. For a broader view on the physical origins of the material dependent effective mass, complying Hamitonian formulations, compatible boundary conditions at heterointerfaces, and their repercussions for electron states in heterostructures see Ref.~\onlinecite{IvchenkoPikusBook1997}. Using a single macroscopic dielectric constant is an approximation which can break down at the atomic scale (due to dielectric screening effects) or near heterointerfaces where disparate materials with different dielectric constants meet, in which cases the Coulomb interaction can be adjusted to incorporate additional physics \cite{penn_wave-number-dependent_1962,fischetti_long-range_2001}}

In addition to being an eigenfunction of $\mathrm{H}$, fermionic $\Psi$ must be anti-symmetric with respect to the interchange of any pair of electrons $ p,q \in 1 \ldots N$:
\begin{equation}
\Psi(\ldots , \tilde r_p , \ldots, \tilde  r_q , \ldots) \, = \,
-\Psi(\ldots , \tilde r_q , \ldots, \tilde  r_p , \ldots)  \, . \nonumber
\end{equation}
This brings a non-trivial spin dependence to the multi-electron eigenspectrum even when $\mathrm{H}$ is spin independent.  

The general methodology that underlies the computational procedure described here is a full configuration interaction (FCI) procedure using orthonormal spatial orbitals. \cite{szabo1996modern} Assuming a set of $M$ orthonormal functions of 
$\mathrm{R}^3$, ``an orbital basis set'', $\{\phi_j\}_{j \,=\, 1}^{M}$, and two orthonormal functions of the spin coordinate, 
$\ket{\uparrow}$ and $\ket{\downarrow}$, one constructs the $2M$ ``spin orbitals''
\begin{equation}
\chi_{2j-1} (\tilde r ) = \phi_j (\bm{r}) \ket{\uparrow},\; \chi_{2j}  (\tilde  r ) = \phi_j (\bm{r}) \ket{\downarrow}  
\end{equation}
for $j = 1, \ldots, {M}$. One can then construct a $N$-electron basis function as a Slater determinant
\begin{equation}
v (\tilde r_1, \, \ldots, \, \tilde r_N)  = \frac{1}{\sqrt{N!} }\det 
\left( 
\begin{array}{ccc}
\chi _{1}(\tilde r_{1}) & \cdots & \chi _{1}(\tilde r_{N}) \\ 
\vdots & \ddots & \vdots \\ 
\chi _{N}(\tilde r_{1}) & \cdots & \chi _{N}(\tilde r_{N})
\end{array}
\right) \nonumber
\end{equation}
where we have selected a distinct subset of $N \leq 2 M$ spin orbitals and relabeled the indices of these selected $\chi$’s to run from $1$ to $N$. By construction, each Slater determinant basis function satisfies the requisite anti-symmetry properties of fermions. Also, since the spin orbitals are orthogonal, the resulting full set of $K = \binom{2M}{N}$ distinct Slater determinant basis functions is an orthonormal set.

From a computational perspective, the FCI procedure consists of using a Rayleigh-Ritz method to approximate eigenfunctions and eigenvalues of $\mathrm{H}$ given by Eq.~(\ref{NparticleS}). One starts with a collection of $K$ orthonormal Slater determinant basis functions $\{v_j\}_{j \,=\, 1}^{K}$ and seeks approximate eigenfunctions that are linear combinations of them. The vectors of coefficients $\bm{c} = (c_1, \ldots, c_K)$ in the expansions of the approximate eigenfunctions are determined by finding the eigenvectors and eigenvalues of the finite dimensional 
linear system $\tilde{\mathrm{H}} \bm{c} = \lambda \bm{c}$  which is the projection of the operator $\mathrm{H}$ onto the subspace formed from the $v_j$'s, i.e., the $(i,j)$ entry of $\tilde{\mathrm{H}}$, $\,   \tilde{\mathrm{H}}_{ij}$, is given by 
$\tilde{\mathrm{H}}_{ij} \, = \, \bra{v_i } \mathrm{H} \ket{v_j}$ . The accuracy of the approximation is improved by expanding the number $M$ of spin orbital basis functions, and, consequently, the number $K$ of Slater determinant basis functions that
can be formed from these spin orbitals. Because each Slater determinant is characterized by the projection of total spin $S_z$, FCI eigenstates of distinct $S_z$ are constructed only using Slater determinants of that projection. This improves the computational efficiency as the subset of Slater determinants for each $S_z$ is reduced.

\subsection{Computational Tasks}

A principal computational task associated with the FCI procedure, and with wave function based methods in general, is the evaluation of inner products involving Slater determinant basis functions,  $\bra{ v_i} \mathrm{H} \ket{v_j}$. These inner products are integrals over the $3N$ spatial coordinates and the $N$ spin coordinates of the $N$-particle wave function. However, as described in Ref.~\onlinecite{szabo1996modern}, since Slater determinant basis functions are linear combinations of products of spin orbitals, the task of evaluating these inner products can be reduced to the task of combining the results of the evaluation of integrals over $\mathrm{R}^{3}$ and over $\mathrm{R}^{6}$. The integrals over $\mathrm{R}^3$ are known as ``one-electron integrals'' and, for real valued orbitals, have the form
\begin{equation}
\mathrm{I}_{ij} = \int_{\mathrm{R}^3} \phi_i (\bm{r}) \, \mathrm{h} \, \phi_j (\bm{r}) \, d \bm{r}
\label{singleIntegrals}
\end{equation}
and must be evaluated for all $j \ge i$, $i = 1, \ldots, M$  pairs of spatial orbital basis functions that are used to create the Slater determinant basis functions. Equations simplify further when the eigenfunctions of the single particle Hamiltonian are chosen as the orbitals.

The integrals over $\R^{6}$ are known as ``two-electron integrals'' and have the form
\begin{equation}
\mathrm{I}_{ijkl} = \alpha
 \int_{\mathrm{R}^6} \frac{ { \phi_i (\bm{r}) \,  \phi_j (\bm{r}) \,  \phi_k (\bm{r}’ ) \, \phi_l (\bm{r}’) } }{
 {| \bm{r} - \bm{r}’ |}} \, d \bm{r} \, d \bm{r}’ \, . \label{doubleIntegrals}
\end{equation}
These integrals are evaluated for all  $j \ge i$, $ i = 1, \ldots, M$ pairs and all $l \ge k$, $k = 1, \ldots, M$ pairs of spatial orbitals. 

A challenge in the construction of grid based numerical methods is that of efficiently and accurately approximating the values of integrals  in Eqs.~ (\ref{singleIntegrals}) and (\ref{doubleIntegrals}) when the spatial orbitals $\phi_i$ are discrete, e.g., they are grid based spatial orbitals. 

For computational efficiency we seek approximations that utilize a computational region that is rectangular and a grid that has a uniform mesh in each direction. For optimal accuracy and efficiency, different mesh spacings $\Delta_\gamma$ in each of the coordinate directions $\gamma=x,y,z$ are often used. With this choice of computational domain and grid, one must design accurate approximations to Eqs.~(\ref{singleIntegrals}) and (\ref{doubleIntegrals}). As explained below, the evaluation of the two-electron integrals in Eq.~(\ref{doubleIntegrals}) can (and should) avoid the direct evaluation of six dimensional integrals. 
 
We use here the standard multi-dimensional trapezoidal method. We assume that the spatial orbital basis functions 
vanish outside the computational domain so that the trapezoidal method will be just a simple uniform weight sum of function
values over all mesh points, e.g.,
\begin{equation}
\int_{\mathrm{R}^3}  f( \bm{r} ) g(\bm{r}) \, d \bm{r} \,\,
\approx  \Delta_x \Delta_y \Delta_z \sum_i \sum_j \sum_k \, f_{ijk} \,  g_{ijk} \,.
\end{equation}
For functions that vanish at the boundaries, the trapezoidal method itself is ``infinite order accurate'', i.e., the convergence improves as a product of positive powers of the mesh spacings, with exponents dependent on the smoothness of the integrand. \cite{atkinson1978introduction}

\subsubsection{One-electron Integrals}
In the construction of the approximation of the one-electron integrals Eq.~(\ref{singleIntegrals}) 
we assume that the spatial orbitals, $\phi_i$, are known, are well resolved by the computational grid, and vanish outside the computational domain (the construction of a set of functions with these properties is described below). The first integral in Eq.~(\ref{singleIntegrals}), due to the kinetic term in $\mathrm{h}$, requires the evaluation of $\phi_i (\bm{r}) \nabla \cdot \bm{\beta}\cdot \nabla \phi_j (\bm{r})$. The difference operators in this term are approximated with standard high order centered finite differences. \footnote{The values outside the computational domain that are required for the evaluation of the finite difference operator at points near the boundaries are taken to be identically zero; these values are consistent with the assumption that the spatial orbitals vanish outside the computational domain.} The integration of the potential term $\phi_i (\bm{r}) \mathrm{U}(\bm{r}) \phi_j (\bm{r})$ in Eq.~(\ref{singleIntegrals}) is straightforward using a trapezoidal method. If the orbital basis functions are smooth functions and vanish outside of the domain, and if the external potential is also smooth, then both kinetic and potential terms in Eq.~(\ref{singleIntegrals}) can be approximated with high accuracy. 

\subsubsection{Two-electron Integrals}
The method for computing two-electron integrals is based on the observation that the evaluation of the six dimensional integral $\mathrm{I}_{ijkl}$ in Eq.~(\ref{doubleIntegrals}) only requires the evaluation of three dimensional operators. As a result, the computational procedure consists of a double loop in which the inner loop evaluates integrals of the form
\begin{equation}
\Phi_{ij}(\bm{r}') = 
\alpha \int_{\R^3} \frac{ { \phi_i (\bm{r})  \phi_j (\bm{r}) } }{ 
{| \bm{r} - \bm{r}' \, | }} \, d \bm{r} \, \label{greenconvolution}
\end{equation}
at the nodes of the computational grid, and an outer loop where the complete integral is evaluated by forming a trapezoidal approximation to the integral
\begin{equation}
\mathrm{I}_{ijkl} = 
\int_{\R^3}  \Phi_{ij}(\bm{r}') \,  \phi_k (\bm{r}' ) \,  \phi_l (\bm{r}')   d \bm{r}' \, . \label{outerIntegral}
\end{equation}
 
The primary computational difficulty in the evaluation of the two-electron integrals is thus in evaluating Eq.~(\ref{greenconvolution}). Since these integrals must be evaluated at each grid 
point of the computational domain, this task is equivalent to determining at all those points the 
solution to the Poisson equation
\begin{equation}
\Delta \Phi_{ij}  = - 4 \pi \alpha \, \phi_i \, \phi_j  \label{infdomainPoisson}
\end{equation}
with ``infinite'' boundary conditions. 

This is a fundamental problem of computational physics and there are many highly efficient algorithms that may be employed. The method one chooses depends upon several factors: the properties of the source functions, available computational hardware, accuracy requirements, etc. For the particular problem presented in this paper the orbital basis functions are not in general strongly localized, and most importantly, the orbital basis functions are not known analytically. The method used here combines two parts: a high order accurate solution of Poisson's equation obtained using discrete sine transforms, and a high order accurate finite difference solution of Laplace's equation that is added so that their sum is a solution of Poisson's equation in an infinite domain. \cite{CANDERinfDomII} The procedure has the advantage that it is a direct method with a computational cost that is $\mathrm{O} (n \log n)$ where $n$ is the total number of grid points, and the bulk of the computational work is carried out by calls to efficient FFT routines. \cite{FFTW05,FFTW3soft} Other methods that are appropriate for uniform grids and utilize FFT routines for efficiency may also be considered.
\cite{James197771,McCorquodale2005,McCorquodale:2007:LCA,Serafini2005481,wang1999efficient,BeylkinGregory2009,VICO2016191,EXL2016629,PhysRevB.73.205119,hejlesen_2013,hejlesen_2016} It is not difficult to imagine problems for which non-uniform grids are used to represent functions, or when the orbital basis functions have high frequency components. For such problems, the collection of methods such as adaptive Fast Multipole Methods (FMM) \cite{langston2011,Malhotra2015} or adaptive fast convolution methods \cite{BeylkinGregory2009} are possible choices.  It is worth mentioning that for the test problem discussed in Sec.~\ref{Sec4}, the accuracy of the solution procedure used to solve Eq.~(\ref{infdomainPoisson}) has been observed to have a significant impact on the precision with which the energies are determined.

\subsection{\label{ssec22} Potential Dependent Orbital Basis} 

The construction of an approximate solution of the $N$-particle Schr{\"o}dinger equation using Slater determinant basis functions requires a specification of a set of spatial orbitals. The use of a uniform grid and spectral approximations dictates that the basis functions be as smooth (differentiable) as possible. With the use of a computational grid, there is great flexibility in selecting basis functions, and in this section we describe a method for creating sets of orbitals that are adapted to the external potential and, assuming a smooth external potential, are also smooth.  

The idea for the construction of these orbitals is an extension of the idea that underlies the use of hydrogenic orbitals as a basis for the classical description of atomic structure. Specifically, hydrogenic orbitals are the eigenfunctions of the single particle Schr{\"o}dinger equation with the external potential being a single nuclear potential. These eigenfunctions are complete and, they, or smooth approximations to them, have proven to be exceptionally useful as basis sets for a variety of \textit{ab initio} computational procedures. The extension of this idea in the context of semiconductor quantum dot modeling consists of using as an orbital basis the eigenfunctions of a single particle operator with the specified external potential. To induce complete localization of the orbital basis to the computational domain, two complementary additions, a barrier potential and a variable kinetic energy coefficient, are incorporated into the single particle operator. Specifically, the single particle operator used to construct the orbital basis has the form
\begin{equation}
\mathrm{h}_b \, = \, -\frac{\hbar ^{2}}{2}  \nabla \cdot  \beta_{b} (\bm{r}) \cdot \nabla + \,
\mathrm{U}_{b} ( \bm{r} ) \, . \label{modifiedBPO}
\end{equation}
Here $\mathrm{\beta}_b$ is an augmented kinetic energy coefficient $\beta$, and $\mathrm{U}_{b}$ is $\mathrm{U}$ augmented by a domain boundary potential. The boundary potential is identically zero in the interior of the computational domain and smoothly transitions to having large positive value over the region of width $b$ near the computational domain boundary. The additional boundary potential cannot be arbitrarily large for numerical reasons \footnote{Using a very large positive value for the potential increases the spectral radius of the operator whose eigenvectors are used to create the basis. If Krylov subspace iterative methods, such as the Rayleigh-Chebyshev method used in this paper,\cite{Anderson2010} are employed, then an increase in spectral radius can substantially reduce the computational efficiency of the iterative procedure for determining the eigenvectors} so it cannot unconditionally localize all the eigenfunctions used for the orbital basis. To ensure localization, $\mathrm{\beta}_b$ is forced to transition from $\mathrm{\beta}$ to zero over the boundary region. Where the kinetic operator is zeroed, the Hamiltonian is completely local and disconnected from the interior of the domain. Hence all the low-energy eigenfunctions either do not encroach on or have vanishing tails outside of the computational domain.

The eigenfunctions are determined numerically as the orthonormal eigenvectors of a high order finite difference approximation to Eq.~(\ref{modifiedBPO}). The computation of the eigenvectors uses the same computational grid as that used for the orbital integrals.  Since some of the eigenvalues and eigenvectors associated with this operator can be degenerate (and more so for the finite dimensional Hamiltonian of the FCI approximation itself), the use of a robust iterative method for determining the eigenpairs was necessary. The particular iterative method used to obtain the results presented here is the Rayleigh-Chebyshev method described in Ref.~\onlinecite{Anderson2010}. We arbitrarily choose the $M$ lowest eigenstates as the basis set, but more elaborate strategies could be considered, possibly improving the convergence of FCI solutions.

\section{Benchmark Problem} \label{Sec3}

In this section we describe the benchmark problem that demonstrates the construction of highly accurate solutions of the $N$-particle Schr{\"o}dinger equations using a completely grid-based FCI procedure.

The problem chosen is an important one for the design of semiconductor qubits. Most implementations of semiconductor qubits use the electron spin as the fundamental quantum system that is manipulated; in some cases, the spin state itself is the qubit, in other cases, two low lying states of two or three electrons form the computational basis. \cite{burkard_semiconductor_2021} For all of these cases, the electrons are confined in quantum dot nanostructures and, depending on how the qubit is defined, the controlled interaction between electron spins in neighboring dots  is an important ingredient for either single or two qubit gates. This interaction usually takes the form of the Heisenberg exchange interaction, $H = J \bm{s}_1  \cdot \bm{s}_2$, where $\bm{s}_1$ and $\bm{s}_2$ are the spin states of electrons in neighboring dots. This interaction arises due to the overlap of the wave functions of the electrons in the two dots and can be controlled by changing the electron confining potential with voltages applied to surface gate electrodes. Faithful simulation of real devices requires detailed modeling of the semiconductor heterostructure potential, strain, and the electrostatic potential created by the gate electrodes through the solution --- possibly self-consistent --- of the Poisson equation with appropriate boundary conditions. Our method is fully compatible with using the results of an auxiliary calculation to provide the electron potential for use in the Schr{\"o}dinger equation. However, such a calculation is not required to demonstrate the accuracy of our method, and so instead we use here an analytical potential representative of those formed in gated semiconductor devices. 

An analytical expression \cite{Davies1995} for the potential generated beneath a row of several square gates of size $2a$ placed next to each other, each held at its own potential, with the surface outside the squares held at $V=0$, is
\begin{equation}
\begin{array}{rll}
\phi_0(x, y, z)    =   -\sum_{i} V_i  \big[ & g (a - x_i + x, a + y, z) + \\
           & g (a - x_i + x, a - y, z)  +  \\
           & g (a + x_i - x, a + y, z)  +  \\
           & g (a + x_i - x, a - y, z) & \big] 
\end{array}
\label{potential2D}
\end{equation}
where
\begin{equation}
g (u, v, z) = \frac{1}{2 \pi} \tan^{-1} \left(\frac{uv}{ z \sqrt{u^2 + v^2 + z^2}} \right) .
\end{equation}
Evaluating Eq.~(\ref{potential2D}) for appropriate choices of the parameters creates the electron confining potential in two of the three directions, $x$ and $y$, but along $z$ another source of confinement is required. In real devices, it is commonly provided by a planar semiconductor heterostructure that utilizes two dissimilar materials, such as Si and a SiGe alloy, to form a one-dimensional (1D) quantum well for electrons. For the calculations described below, we assume the heteropotential of a $5$~nm wide square well with a depth of $238$~meV, typical numbers for a Si/SiGe heterostructure; it spans $15$~nm in total to allow for some wave function penetration into a $5$~nm barrier regions on either side of the well. The full 3D confining potential is then constructed here as a product of this 1D square well potential and the 2D potential, described by Eq.~(\ref{potential2D}) at a specific depth $\bar z$ corresponding to the middle of the quantum well.  Other constructions are certainly possible, but found to be inconsequential for the present analysis. 

We focus below on a case of two interacting electrons in a double quantum dot geometry. With only three gates total, we can shape the electrostatic confinement to form potential minima under the outer gates with a tunable barrier between them by varying gate voltages.

\begin{figure}
 \includegraphics[width=\columnwidth]{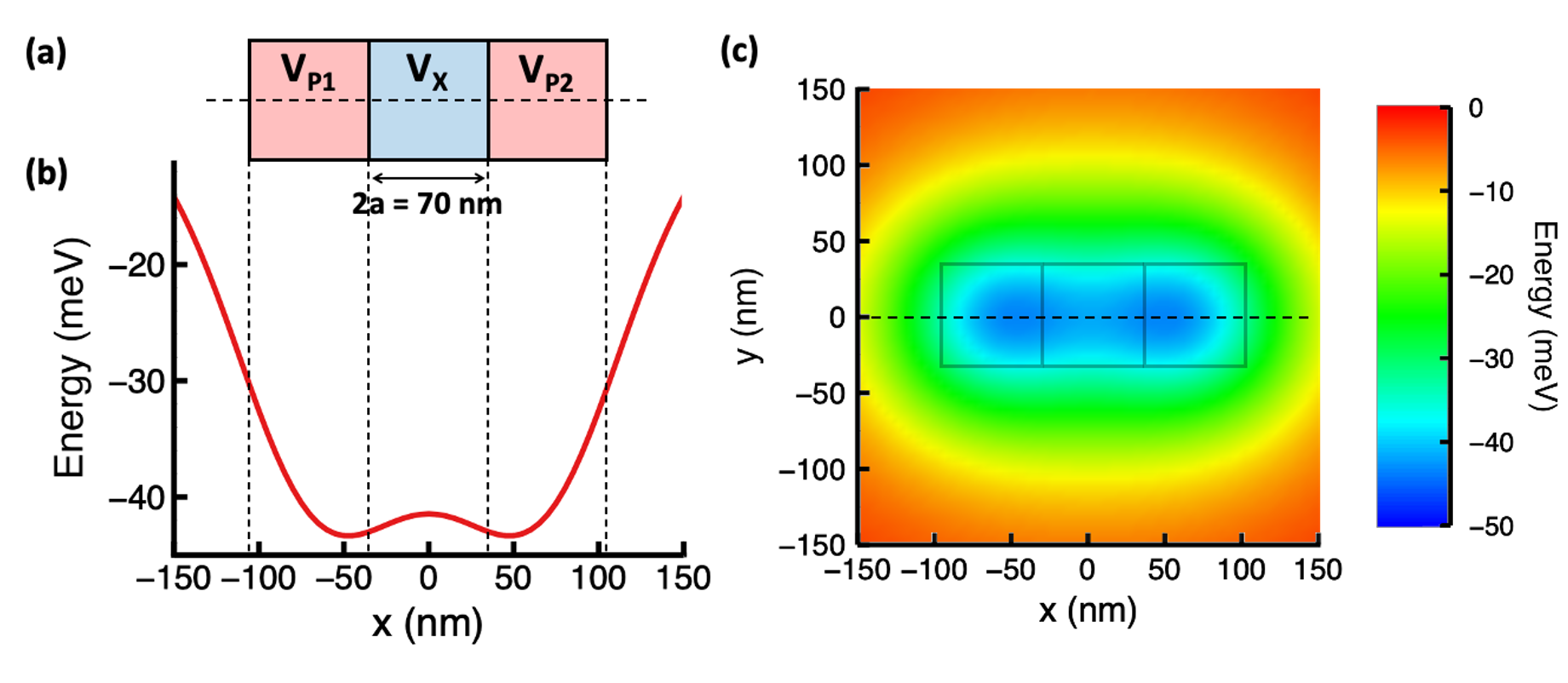}
\caption{(a) Schematic of the square gates used to generate the in-plane confining potential. (b) 1D slice of the model 2D confining potential taken at the midpoint of the gates as indicated by the dashed line in panel (a). 
(c) The full 2D potential with the color scale indicating its depth.
}
\label{fig-1} 
\end{figure}

Figure~\ref{fig-1} describes the model three-gate potential generated using Eq.~(\ref{potential2D}) by setting $2a=\bar z=70$~nm --- typical for quantum dot qubit devices, $x_i = -2a$, $0$, and $2a$, and the gate voltages $V_i = V_{P1}$, $V_X$, and $V_{P2}$ to $0.25$, $0.10$, and $0.25$~V, respectively. Panel~(a) is a schematic of the three adjacent square gates.The dashed line cutting horizontally through the gates is at their midpoint and panel~(b) shows a 1D slice of the potential taken along that cut. For these applied voltages, we see that two potential minima are formed under the outer gates with a small barrier under the middle gate, consistent with the outer gates being held at a slightly more positive bias than the middle gate. This is the configuration that quantum dot qubits typically operate in, where the barrier between the electrons is sufficient to keep them separated but still interacting through overlap of the tails of their wave functions. This interaction can be controlled in one of two ways --- by modulating the height of the potential barrier through changing the voltage on the middle gate, $V_X$, or by changing the depth of one potential minimum relative to another, known as detuning, through asymmetrically biasing the outer gates, $V_{P1}$ and $V_{P2}$. We will examine both of these modes of operation in the next section. Figure \ref{fig-1}(c) shows the full 2D potential with the colors indicating its depth. The dashed line again corresponds to the location of the 1D cut shown in panel (b). The lateral dimensions of the computational domain are $300$~nm for this case and for all the results below. While the potential has not fully relaxed to zero at the edge of the domain for these dimensions, note that its value at the edge is at least  $25$~meV above the bottom of the two potential minima. This is a rather deep confining potential for electrons in Si (with $m_x=m_y=0.19m_0$, $m_z=0.92m_0$, where $m_0$ is the free electron mass, and $\epsilon = 11.7\epsilon_0$ where $\epsilon_0$ is the vacuum permittivity), and the ground state wave function of the single particle Schr{\"o}dinger equation is effectively zero (below machine precision) well before reaching the edge of the domain. The modifications of the single particle Hamiltonian near domain boundaries, introduced in Eq.~(\ref{modifiedBPO}), ensure that all basis orbitals, including highly excited ones, vanish at the edge of the domain.

\section{Computational Results} \label{Sec4}

We first demonstrate the accuracy of our method by computing the exchange interaction in two-electron systems using the confining potential of the benchmark problem. The test consists of modulating the applied voltages so that the strength of the exchange interaction $J$ varies from the peV to meV scales. For a two-electron system, the interpretation is straightforward in terms of total spin, and $J$ is equal to the difference between energies of the lowest triplet and singlet states. Since the overall wave function must be antisymmetric as for any fermion system, the singlet (triplet) spatial part of the wavefunction is symmetric (antisymmetric) with respect to electron interchange, facilitating state attribution. This energy difference can get extremely small if the electrons are separated by a large potential barrier, meaning their spins become decoupled. With this background, we now can lay out the computational tasks (in fact, applicable to any number of electrons $N$) that must be performed for each bias:

\begin{enumerate}[(i)]
 \item Compute the lowest $M$ eigenfunctions of the single particle operator in Eq.~(\ref{modifiedBPO}).              
 \item Evaluate all one- and two-electron integrals associated with these $M$ orbitals.
 \item For each distinct value of total spin projection $S_z$, construct the discrete $N$-electron Hamiltonian matrix from the corresponding subset of Slater determinants using the one- and two-electron integrals.
 \item Determine the eigenvalues and eigenvectors of the resulting Hamiltonian matrix. 
 \item \emph{For two-electron exchange:} Take the difference between the lowest triplet and singlet to obtain $J$.
\end{enumerate}

There are two types of errors associated with this computational procedure. The first type, the``numerical discretization error'', 
is the error associated with implementing discrete approximations for the differential and integral operators whose evaluation is required to set up and solve the equations. This includes also the error from approximating the orbitals numerically. The second type is the ``orbital basis set error'' due to approximating the desired  $N$-particle eigenfunctions with a linear combination of Slater determinants formed from a \emph{finite} set of spatial orbitals. 

The potential associated with this benchmark problem is discontinuous in the $z$-direction at the quantum well interfaces, and, as a consequence, the order of accuracy of the integral and differential operators in the $z$-direction is reduced to second order (as long as the quantum well interfaces are coincident with computational grid planes). Since the order of accuracy in the in-plane directions is not affected, consistently accurate solutions can be obtained simply by using a more refined mesh in the out-of-plane direction. All the results presented below were obtained using 
a uniform mesh with $5$~nm spacing in the $x$ and $y$ directions and $0.25$~nm in the $z$ direction. \footnote{Using the domain dimensions stated above, $300$~nm in the in-plane directions and $15$~nm along $z$, this results in a total of $n=226981$ points that define the computational grid. This is therefore the size of the system that defines the discretized single particle Hamiltonian and associated orbital basis states which are integrated over to obtain the orbital matrix elements.} The mesh spacings were chosen sufficiently small so that the dominant error in the exchange energy was due to orbital basis set size. 

The orbital basis set is adaptive in the sense that every distinct bias (and hence device potential) can lead to a distinct set of orbital states that are used in the corresponding FCI calculation. Empirically we find that a relatively modest number of orbitals is sufficient for accurate calculation of the exchange energy.

\begin{figure}
 \includegraphics[width=\columnwidth]{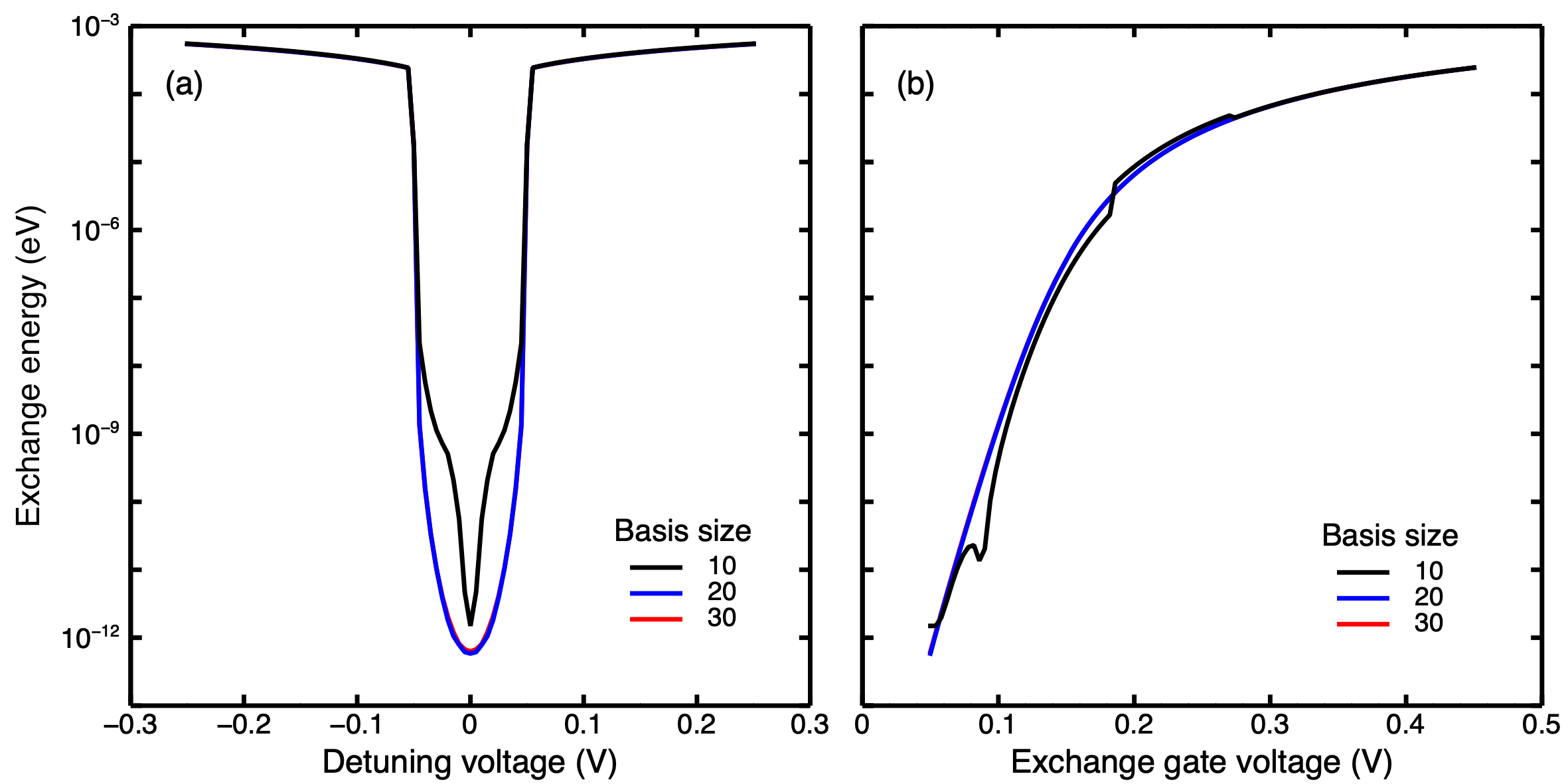}
\caption{Exchange energy calculated for different basis set sizes using the model potential described in the text. (a) $J$ as a function of detuning voltage $V_{P2} - V_{P1}$ for exchange gate bias $V_X=0.05$~V; at zero detuning, $V_{P1}=V_{P2}=0.25$~V. (b) $J$ versus $V_X$ at zero detuning with $V_{P1}=V_{P2}=0.25$~V.}
\label{fig:JvsVs}
\end{figure}

Figure \ref{fig:JvsVs}(a) shows the calculated exchange energy as a function of detuning voltage at a fixed exchange gate bias $V_X = 0.05$~V. The detuning voltage is defined as the difference between $V_{P2}$ and $V_{P1}$. Although this is just a model potential, these detunings are in the range of what is applied to real devices to achieve practical values of exchange energy. The data shown is for three different basis set sizes, $10$, $20$, and $30$. The basis functions themselves are the eigenstates of the single particle Hamiltonian as described in Eq.~(\ref{modifiedBPO}). For large values of detuning, $J$ is on the order of a few hundred $\mu$eV and varies weakly with bias, representing the situation when the two electrons collapse into the same dot. The suppression of $J$ at smaller values of detuning is due to the two electrons becoming spatially separated into the left and right dots with a potential barrier between them. The abruptness of this transition depends on the exchange gate voltage, $V_X$. For larger $V_X$, the interdot barrier is reduced, the minimum exchange increases and this transition is more gradual. We chose the case in Fig.~\ref{fig:JvsVs}(a) because it highlights that we are able to achieve well-converged energies even at the peV scale. Indeed, with only $20$ basis functions, the exchange energy is already well-converged and increasing the basis set size does not significantly change the results on an absolute scale. Good convergence over more than $8$ orders of magnitude in $J$ is due to using an adaptive basis which is the hallmark of our approach. While convergence over such a large range may not be necessary in other cases, it is actually quite important for modeling semiconductor qubits. Controlling the exchange interaction over this range is essential to the operation of these devices. For example, knowing that the minimum value of $J$ (the so-called residual exchange) for a particular set of voltages is, say, 10 peV rather than 1 neV can be the difference between a functional and a faulty qubit tune-up (or even design), as large residual exchange causes significant errors in qubit operation that are not easily corrected.

Figure \ref{fig:JvsVs}(b) shows $J$ as a function of exchange gate bias $V_X$ at zero detuning with $V_{P1}=V_{P2}=0.25$ V; this modulation keeps the potential symmetric along $x$ and the dots singly occupied, resulting in a smoother variation of $J$ compared to the detuning case. The exchange energy is approximately exponential in  $V_X$ at lower voltages where the potential barrier is large and only the tails of the electron wave functions are interacting. The simplest theories for barrier tunneling explain this behavior \cite{Bhattacharya1982} and are confirmed here in a more complex potential. At larger $V_X$ values, the potential barrier is nearly zero; in this regime, the two electrons end effectively in one large, shallow potential well and a saturation of $J$ is expected. As was already observed for the detuning case, the calculation converges quickly with basis set size and is well converged across the whole range of $J$ values.

\begin{figure}
 \includegraphics[width=\columnwidth]{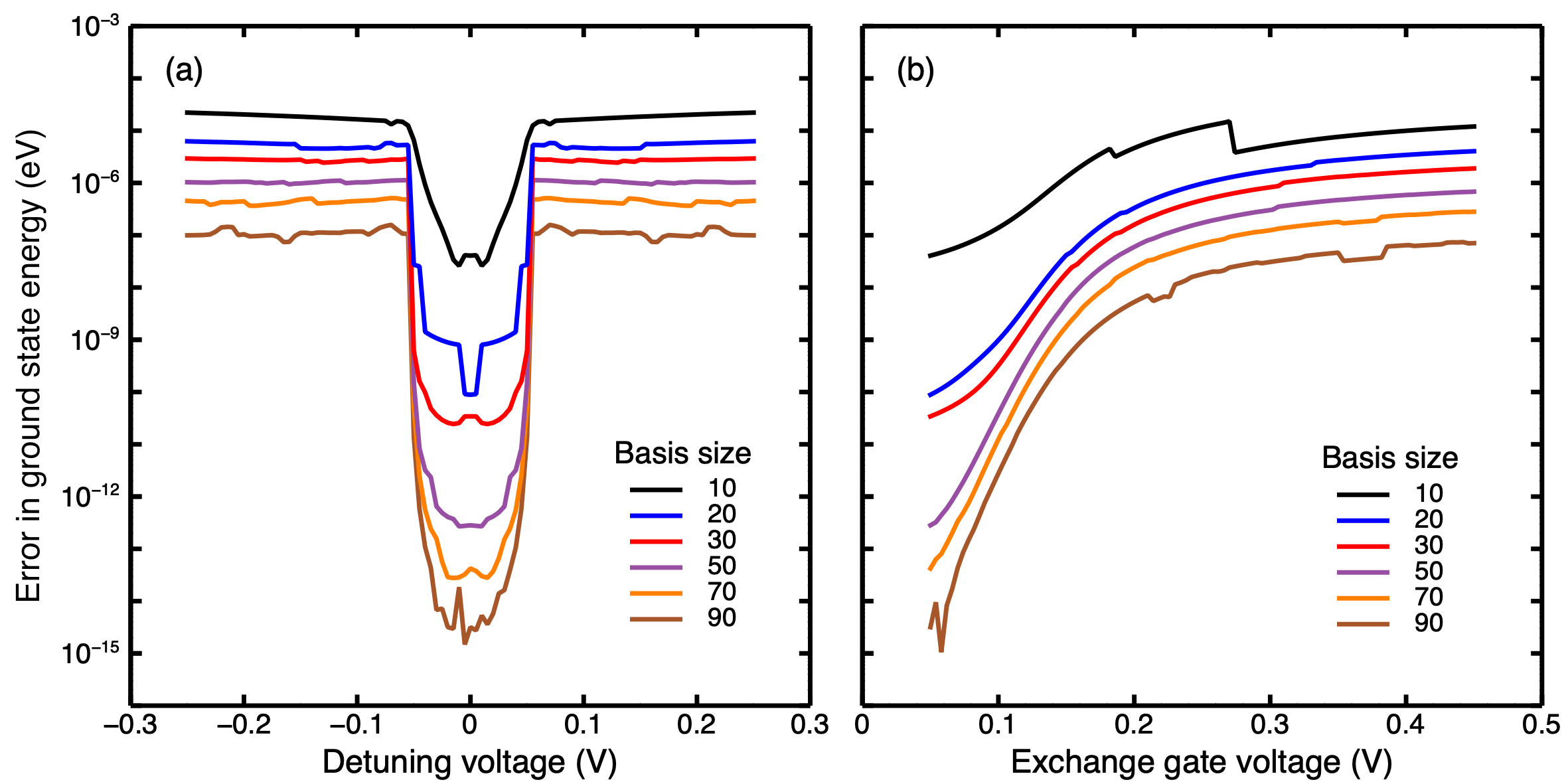}
\caption{Absolute error in the ground state energy $E_0$ as the (a) detuning and (b) exchange gate voltages are varied. For each reported basis set size, it is estimated as a difference in $E_0$ with value calculated using basis set size of $100$.}
\label{fig:E0error}
\end{figure}

\begin{figure}
 \includegraphics[width=\columnwidth]{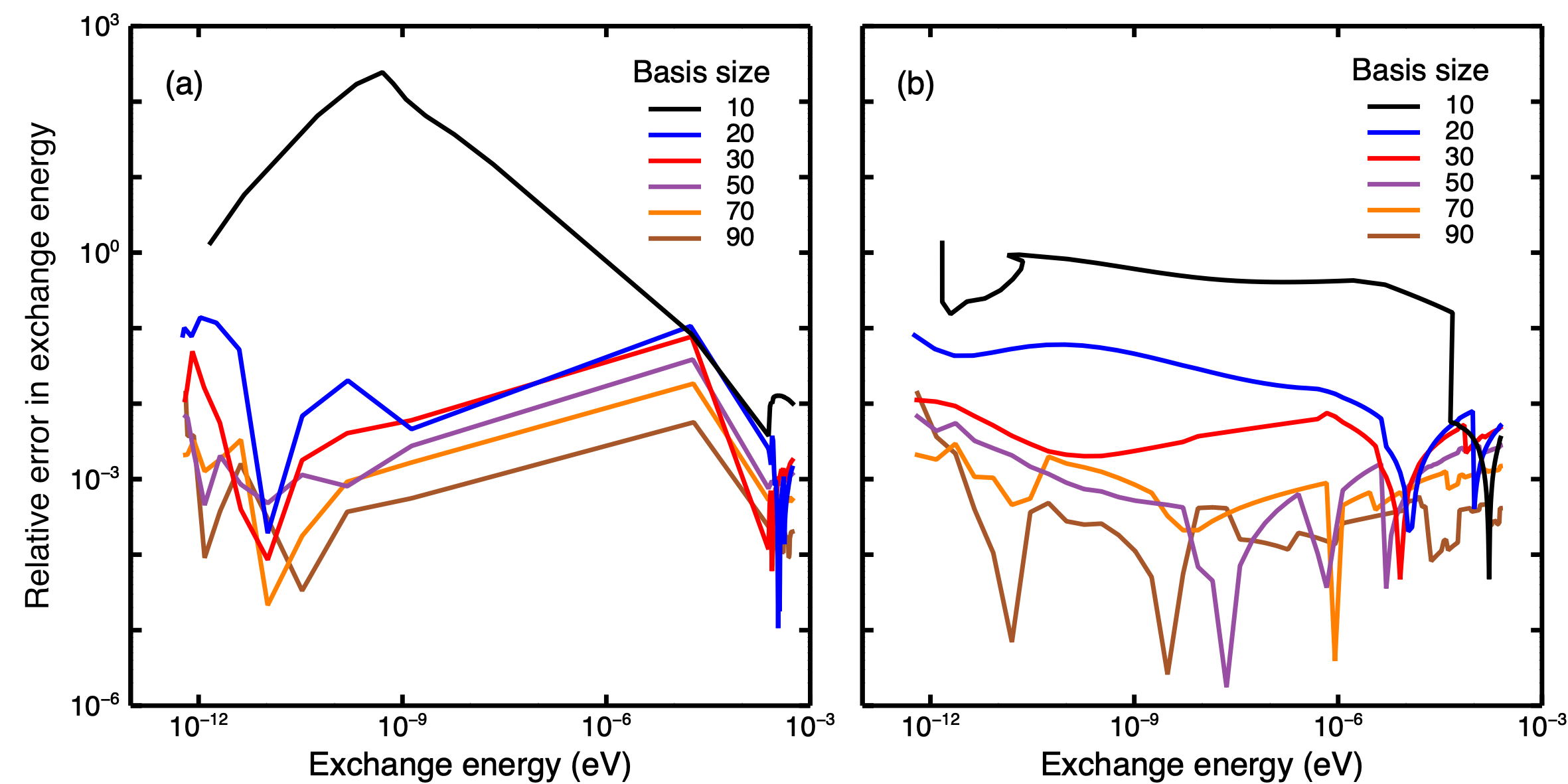}
\caption{Relative error in exchange energy $J$ for (a) detuning and (b) exchange gate voltage sweeps. Shown for several basis set sizes as a function of $J$ calculated using a basis set size of $100$.}
\label{fig:Jerror}
\end{figure}

What we have demonstrated so far is convergence with respect to basis set size on an absolute scale, meaning that we can compute the exchange energy over a very large range and be confident that the calculation is not off by a significant factor even for extremely small values of $J$ provided that the basis set is sufficiently large, e.g., greater than~$20$. The plots shown give very little indication of the relative error in $J$, however. 

The exchange energy is determined by the difference of the two lowest energy values, and we turn next to estimating the accuracy with which the energies themselves are being computed. Exact values for the energies of the system are not known, and so an estimate of the error is obtained by comparing the results of computations with orbital basis sets of sizes from $10$ to $90$ with those obtained with the largest basis set of size~$100$. Figure~\ref{fig:E0error}(a) and~(b) show the values of the estimated error for the overall ground state (which is a singlet) energy as the detuning and exchange biases are varied. For every bias, the energy values converge monotonically as the basis set sizes increases. The behavior of the estimated error of the first excited state (which is the ground triplet) was similar and so is not shown.  A prominent feature in these results is that the convergence behavior is not uniform with respect to bias voltages. In particular, as the biases are varied so $J$ decreases, the estimated accuracy of the energies improves. This behavior can be explained by the observation that as $J$ decreases, the configuration tends to one whose electronic structure is associated with two spatially separated electrons interacting primarily electrostatically.  Since the orbital basis used consists of eigenfunctions of the single particle operator, one expects more rapid convergence. The second feature, abrupt jumps in the estimated errors, is because we elected to use the $M$ \emph{lowest} eigenstates as the basis set. This can lead to inclusion or exclusion of an important orbital when its energy dips or rises relative to other states, and is most extreme at smallest $M$.

Figure \ref{fig:Jerror}(a) and~(b) shows the estimated relative error in $J$ for the detuning and exchange gate voltage sweeps for basis sets of size between $10$ and $90$.  The error is defined as the absolute value of the difference between $J$ extracted for a given basis set size and a basis size of $100$, divided by the value of $J$ for basis size of $100$ (which is also used as the abscissa value).  This data shows that, aside from the basis size of 10, which is obviously not well converged as shown in the previous plots, the relative error is generally within a few percent, and, using larger basis sets, can readily be pushed below one percent over the entire range of $J$. Since $J$ is computed by subtracting two nearly equal energy values, the constraints of finite precision arithmetic and computed eigenvalue accuracy set a floor on the resolvable relative error in $J$. Thus, for the smallest values of $J$ the non-monotonic trend of the relative errors as the basis size increases is not unexpected.

\section{Conclusion}
This paper describes a collection of discretization techniques that can be combined to create a real space, grid based method for the determination of energies and wave functions of the $N$-particle Schr{\"o}dinger equation used to approximate electrostatically confined electron states in semiconductor quantum dots. In this method, both high order finite difference and spectral approximations of differential and integral operators are used to obtain high accuracy. High computational efficiency for the evaluation of the two-electron integrals is obtained through the use of high performance fast Fourier transform 
routines. \cite{FFTW05,FFTW3soft} The use of completely grid-based approximations facilitates use of numerically determined orbital basis functions. In particular, one can construct and utilize orthonormal orbital basis sets that
consist of eigenfunctions of a single particle operator with arbitrarily defined external potentials. This capability allows for the creation of adaptive orbital basis sets for problems in which the external potential varies greatly. 

A benchmark problem is presented that incorporates an analytically described confining potential representative of potentials produced in electrostatically gated devices. The computational results demonstrate that when our method is applied to the benchmark problem, the energy differences between the ground singlet and triplet two-electron states are accurate to within a percent over a wide range of model parameters and resulting exchange values. 

While the use of the discretization techniques has been demonstrated for the implementation of a FCI procedure for electrons in a semiconductor, the same techniques can certainly be utilized in other problems, for example standard molecular modeling, \cite{Anderson2018} or implementations of other \textit{ab initio} procedures such as Hartree-Fock or density functional theory. \cite{Martin2004} They may also find application in the construction of methods that necessitate the use of real-space grids, in particular, methods that combine different types of quantum mechanical approximations
in different regions of physical space. 

\begin{acknowledgments}
We would like to acknowledge Ekmel Ercan for careful reading of the manuscript. This work was supported in part by the DARPA Quantum Information Science and Technology (QuIST) Program (ARO DAAD-19-01-C-0077).
\end{acknowledgments}

\section*{Author declarations}
\subsection*{Conflict of interest}

The authors have no conflicts to disclose.

\section*{Data Availability}

Data available on request from the authors.

\section*{References}

\bibliography{GridBasedDeviceQM}

\end{document}